\documentclass[prb,twocolumn,showkeys,showpacs]{revtex4} 
\usepackage{epsfig}
\newcommand{\mmt}{Hg$_{\scriptstyle 0.98}$Mn$_{\scriptstyle 0.02}$Te\,}
\newcommand{\mmxt}{Hg$_{\scriptstyle 1-x}$Mn$_{\scriptstyle x}$Te\,}
\newcommand{\cmt}{Hg$_{\scriptstyle 0.3}$Cd$_{\scriptstyle 0.7}$Te\,}
\newcommand{\mct}{Hg$_{\scriptstyle 1-x}$Cd$_{\scriptstyle x}$Te\,}
\newcommand{\czt}{Cd$_{\scriptstyle 0.96}$Zn$_{\scriptstyle 0.04}$Te\,}

\DeclareMathAlphabet{\mathitb}{OT1}{cmr}{bx}{sl}
\begin{document}
\title{Interplay of Rashba, Zeeman and Landau splitting in a
magnetic two dimensional electron gas}
\date{\today}
\author{Y. S. Gui$^*$}
\author{C.~R.~Becker}
\email{becker@physik.uni-wuerzburg.de}
\author{J.~Liu}
\author{V.~Daumer}
\author{V.~Hock}
\author{H.~Buhmann}
\author{L.~W.~Molenkamp}
\affiliation{Physikalisches Institut(EP 3), Universit\"{a}t
W\"{u}rzburg, Am Hubland, 97074 W\"{u}rzburg, Germany}
\begin{abstract}
The transport properties of a magnetic two dimensional electron gas
consisting of a modulation doped
$n$ type \mmt/\cmt quantum well, QW,
have been investigated.
By analyzing the Shubnikov-de Haas oscillations and the node
positions of their beating patterns, we have been able to separate the
gate voltage dependent
Rashba spin-orbit splitting from the temperature dependent giant Zeeman
splitting. It has been experimentally demonstrated that the
Rashba spin-orbit splitting is
larger than or comparable to the $sp-d$ exchange interaction
induced giant Zeeman splitting in this magnetic 2DEG even at moderately
high magnetic fields.

\end{abstract}
\maketitle


The study of a two dimensional electron gas, 2DEG, in a system
with additional interactions has lead to interesting new physics.
One example is the Rashba spin-orbit, s-o,
interaction\cite{Rashba} which has been the subject of numerous
investigations of III-V heterostructures.\cite{Das,Das90,Engels}
More recently, Rowe {\it et al.}\cite{Rowe} showed that
magneto-intersubband scattering, MIS, may cause similar beating
patterns in the magneto-resistance of InAs. Zhang~\cite{Zha02}
{\it et al.} have demonstrated that MIS is not present in narrow
gap II-VI materials, {\it i.e.} HgTe based quantum wells, due to
the strong non-parabolicity of the conduction band. The Rashba s-o
coupling is strong in such HgTe QWs, which exhibit an inverted
band structure when the well width is greater than approximately 6
nm. The exceptionally large zero field Rashba splitting of up to
17 meV observed in this material is due to the heavy hole
character of the first conduction band, which is usually labelled
as $H1$ (see Ref.~\onlinecite{Zha01} for details).

Another example of novel 2DEG physics is the recently introduced
magnetic heterostructure\cite{Smorch97} in which magnetic ions
(usually Mn ions) are exchange coupled to the 2DEG. These structures
have previously been employed to investigate spin
interactions,\cite{Crooker} spin dependent transport and
localization,\cite{Smorch97}  quantum phase
transitions,\cite{Smorch98} as well as to measure the
magnetization of a magnetic 2DEG.\cite{Harris}

So far, magneto-transport studies of a magnetic two-dimensional electron
gas in narrow-gap heterostructures have not been
reported. Such structures offer the intriguing possibility to
probe the interplay of Rashba, Zeeman and Landau effects, which
all may lead to level splittings of comparable magnitude: the 2D
electrons are coupled to the local moments of magnetic ions via a
ferromagnetic $sp-d$ exchange interaction, resulting in spin
splitting energies, $\Delta E_S$, of tens of meV, which are
comparable to or larger than the Landau level splitting, $\hbar
\omega_c$, and finally, as discussed above, the Rashba  s-o
interaction is of a similar magnitude in type III
heterostructures.\cite{Zha01}

In this letter, we report on an investigation of the gate voltage
and temperature dependent Shubnikov-de Haas, SdH, oscillations in
$n$ type \mmt{}/\cmt{} QWs. The observed beating patterns in SdH
oscillations show the effects of a combination of Rashba s-o
interaction and $sp-d$ exchange interaction. The resulting total
spin splitting energy is consistent with the calculated Rashba s-o
splitting energy and a temperature dependent Zeeman component. The
Rashba splitting energy can be varied by a factor of 3.5 in the
low carrier concentration range by means of the gate voltage. The
population difference between the two subbands has been determined
from a fast Fourier transformation, FFT, of the SdH oscillations.
Self-consistent Hartree calculations\cite{Pfeuffer,Zha01} result
in theoretical values for the population difference between these
two subbands which are in good agreement with experiment.


Fully strained $n$ type \mmt{}/\cmt{} QWs were grown by molecular beam
epitaxy, MBE, on \czt(001) substrates in a Riber 2300
MBE system as has been described elsewhere.\cite{Beck00} Results for
the QW discussed in detail here, Q1697, and similar QWs are comparable.
Q1697 was modulation doped symmetrically on both sides of the \mmt{} QW
using CdI$_2$ as a doping material. The \mmt{} well width is 12.2 nm
and the \cmt{} barriers are composed of a 5.5~nm thick
spacer and a 9~nm thick doped layer. Standard Hall bars were
fabricated by wet chemical etching. A 200~nm thick Al$_2$O$_3$
film was deposited on top of the structure, which serves as an
insulating layer. Finally Al was evaporated to form a metallic
gate electrode. Ohmic indium contacts were fabricated by thermal
bonding.

Magneto-transport measurements were carried out in a $^3$He
cryostat using dc techniques with currents of 1~$\mu$A in magnetic
fields ranging up to 7~T and temperatures down to 0.38~K. The
carrier concentration, $n_{H1}$, and the mobility of the first
conduction band were determined to be
$2.4~\times~10^{12}$~cm$^{-2}$ and $5.2~\times~10^4$~cm$^2$/(Vs)
at 4.2~K for zero gate voltage.\cite{E2}


\begin{figure}[htb]
\epsfig{figure=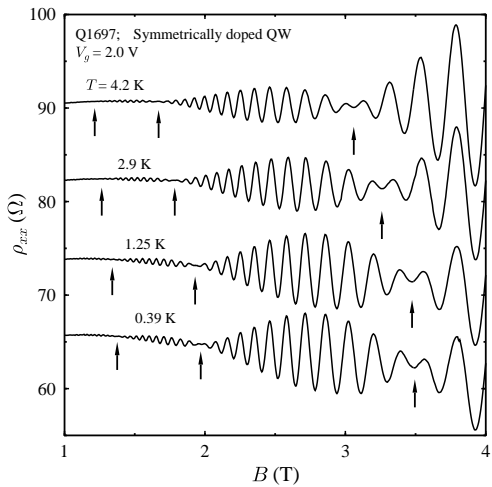,width=0.8\linewidth}
\caption{\footnotesize {SdH oscillations of the symmetrically
modulation doped QW, Q1697, as a function of temperature. 
Node positions in the beating patterns are indicated with arrows. 
All traces are shifted vertically for clarity.}}
\label{q1697rxx-t}
\end{figure}

\begin{figure}[htb]
\epsfig{figure=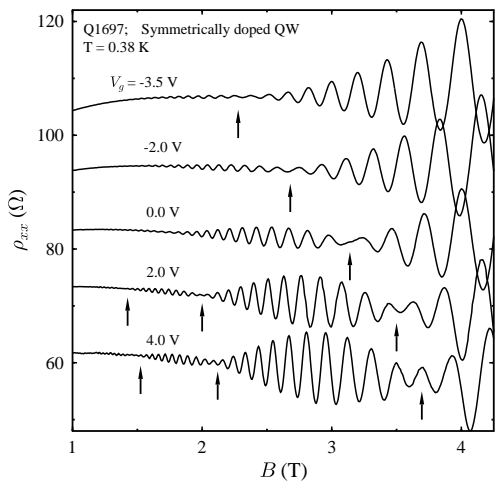,width=0.8\linewidth}
\caption{\footnotesize {SdH oscillations of the symmetrically
modulation doped QW, Q1697, as a function of gate voltage. 
Node positions in the beating patterns are indicated with arrows.
All traces are shifted vertically for clarity.}}
\label{q1697rxx-vg}
\end{figure}

Figs.~\ref{q1697rxx-t} and \ref{q1697rxx-vg} show the temperature
dependence and gate voltage dependence of SdH oscillations for
Q1697. Beating patterns are observed in the SdH oscillations
because of the existence of two closely spaced frequency
components with similar amplitudes due to level splitting.  These
beating patterns show characteristics of both Zeeman spin
splitting (dependence on temperature)  and Rashba s-o splitting
(dependence on gate voltage). To our knowledge, this is the first
observation of the occurrence of both effects in a single sample.
This is the main result of this letter. We analyze our
observations in the following section in order to extract the 
relative magnitudes of these effects.

It is well known that a splitting of the Landau levels leads to a
modulation of the SdH amplitude according to
\begin{equation}
\label{cos}
A \propto \cos(\pi \nu)
\end{equation}
where $\nu$ is given by
\begin{equation}
\label{nu}
\nu = \frac{\delta}{\hbar \omega_c}{\rm ,}
\end{equation}
$\hbar \omega_c$ is the Landau level separation energy and $\delta$
is the energy splitting of each Landau
levels.\cite{Das} Nodes in the beating pattern in the SdH
oscillations will occur at half-integer values of $\nu$; $\pm
0.5$, $\pm 1.5$, etc.; where A is zero. Three nodes are observed
at higher charge carrier concentrations, whereas their number
decreases at lower concetrations. Experimentally, we find
that the first node corresponds to $\nu = 1.5$, and the
successively lower nodes occur at $\nu = 2.5$ and 3.5.

The large temperature shift of the nodes observed in Q1697 and
shown in Fig~\ref{q1697rxx-t}, is caused by the strong $sp-d$
exchange interaction between the conduction electrons and the Mn
ion spins, as discussed by Gui {\it et al.}\cite{Gui} for bulk
\mmxt{}. The temperature dependence stems from the reduction in
magnetization of the Mn ions with increasing temperature.
Phenomenologically, the effective $g^*$ factor in dilute magnetic
semiocnductors can be expressed as\cite{Smorch97,Brandt}
\begin{equation}
\label{gfactor}
g^* = g_0 - \frac{(\Delta E)_{max}}{\mu_B B}
B_{5/2}\bigg [\frac{5g_{Mn}\mu_B B}{2k_B (T+T_{\circ})}
\bigg ]
\end{equation}
where $g_{Mn} = - 2$ is the $g$ factor for Mn, $B_{5/2}(x)$ is the
Brillouin function for a spin of $S = 5/2$, which has been
empirically modified by using a rescaled temperature, $T + T_0$,
in order to account for antiferromagnetic spin-spin interaction,
and $(\Delta E)_{max}$ is the saturated spin splitting energy
caused by the $sp-d$ exchange interaction. $g_0$ is the $g$ factor
for a HgTe QW without the presence of Mn,\cite{Zha01} i.e., $g_0 =
-20$.

\begin{figure}[htb]
\epsfig{figure=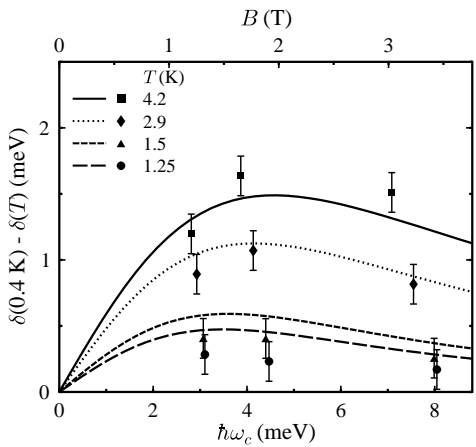,width=0.8\linewidth}
\caption{\footnotesize {Experimental values of the difference in
level splitting energies, $\delta$, between 0.38 K and the
temperature $T$ versus the Landau level splitting energy for the
symmetrically modulation doped Q1697 with $V_g = 1.0$~V. The
curves are the results of a least square fit of values of
$g^*\mu_B B$ by means of Eq.~\ref{gfactor}}} 
\label{zeeman}
\end{figure}

An experimental estimate for $(\Delta E)_{max}$ can be obtained
from the results displayed in Fig.~\ref{zeeman}, where the
differences in experimental level splitting energies, $\delta$,
between 0.38 K and temperature $T$ are plotted versus the Landau
level splitting energy, $\hbar \omega_c = \hbar (e/m^*)B$. The
electron effective mass employed here has been determined by means
of self-consistent Hartree calculations.\cite{Pfeuffer98} The
Hartree calculations are based on a $8~\times~8 \,{\mathitb
k\cdot}{\mathitb p}$ band structure model\cite{Pfeuffer,Zha01}
including all second order terms.
The energy gap of \mmt{} has been taken into consideration;
however, $sp-d$ interaction of the Mn spins has been neglected.
The inherent bulk inversion asymmetry of \mmt{} and \cmt{} has
also been neglected, since this effect  has been shown to be very
small in \mct{}.\cite{Weiler} We find $m^* = 0.047$ to $0.051 \,
m_0$ for the carrier concentrations observed in the experiments,
i.e., $n_{H1} = 2.2$ to $2.6 \times 10^{12}$ cm$^{-2}$ for $V_g$
between $-3.75$ and $+4.75$~V. The curves in Fig.~\ref{zeeman} are
results of least square fits of Eq.~\ref{gfactor} at the
corresponding temperatures. The agreement is reasonable and
results in $(\Delta E)_{max} = 4.3 \pm 0.5$~meV and $T_0 = 2.6 \pm
0.5$~K. In principle, these parameters depend only on the Mn
composition.

However, the giant Zeeman effect can not explain why
the observed nodes also shift with gate voltage, i.e.,
the asymmetry of the QW structure, as can be seen in Fig.~\ref{q1697rxx-vg}.
This behavior is typical for level splitting due to the Rashba s-o
component, as was discussed for non-magnetic HgTe quantum wells 
by Zhang {\it et al.}\cite{Zha01}
A fast Fourier transformation, FFT, of the SdH oscillations as a
function of $1/B$ has been used to determine the carrier concentration
of the
spin split $H1$ subbands.
A double peak structure is
clearly resolved in the FFT spectra (not shown here) of the SdH
oscillations for different
temperatures and gate voltages  shown in Figs.~\ref{q1697rxx-t} and
\ref{q1697rxx-vg} for Q1697.
The  node at the highest field shifts from 2.25 T
($V_g = -3.75$~V)
to 3.72 T
($V_g = 4.75$~V)
while the carrier concentration for the
first conduction subband, $H1$, changes only from $2.24$ to
$2.65~\times~10^{12}$~cm$^{-2}$ and the total carrier concentration
from $2.73$ to $3.56~\times~10^{12}$ cm$^{-2}$.

\begin{figure}[htb]
\vspace{-2mm}
\epsfig{figure=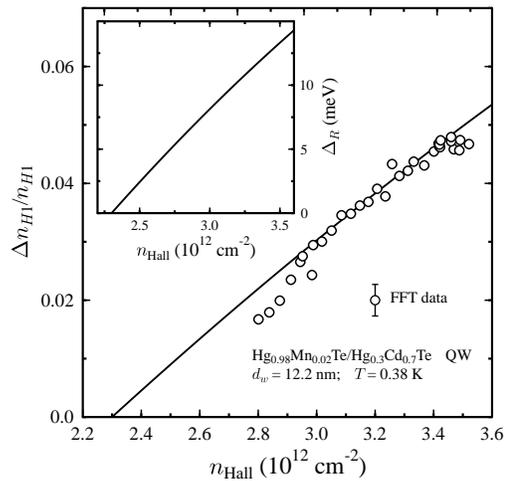,width=0.8\linewidth}
\vspace{-5mm} \caption{\footnotesize {Relative experimental and 
theoretical population differences, $\Delta n_{H1}/n_{H1}$, 
(circles and line) between the two spin states of the $H1$ subband as 
a function of the total carrier concentration, $n_{\rm Hall}$.
The line in the inset represents the corresponding Rashba 
splitting energies, $\Delta_R$, which are extracted 
from the values of $\Delta n_{H1}/n_{H1}$ 
by means of the 8-band 
${\mathitb k\cdot}{\mathitb p}$ model described in the text.}} 
\label{delta-n-new}
\end{figure}

Fig.~\ref{delta-n-new}  shows the relative experimental differences in
population between the two sub-levels of the first conduction
subband, $\Delta n_{H1}/n_{H1}$,
together with the results of self-consistent Hartree calculations
versus the total carrier concentration. In this case, the
calculations are employed to extract the Rashba s-o splitting energy
for the different carrier concentrations.
The calculated Rashba s-o splitting energies (for B=0) are also
plotted in the inset of Fig.~\ref{delta-n-new}. As demonstrated
below, their magnitude of up to 13 meV is greater than that of
Zeeman and Landau level splitting for magnetic fields up to 4 to
5~T.

\begin{figure}[htb]
\epsfig{figure=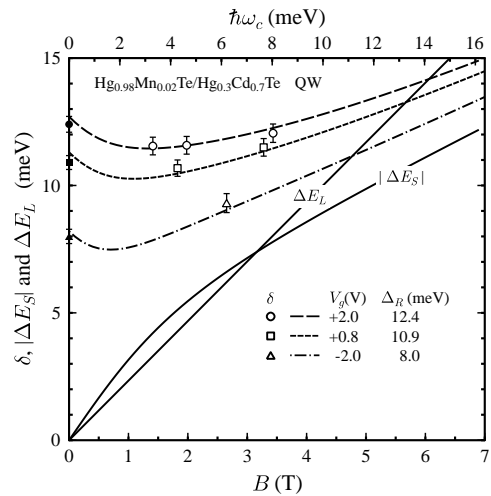,width=0.8\linewidth}
\caption{\footnotesize {Experimental values (empty symbols) for
the total level splitting energy, $\delta$,
versus $B$ and the Landau level splitting, $\hbar \omega_c$
for the symmetrically modulation doped Q1697 for three gate voltages
at $T = 0.38$~ K. The corresponding calculated Rashba s-o 
splitting energies are indicated by filled
symbols. The curves are the results for $\delta$ via Eq.~\ref{delta}
at these three gate voltages, for
$\Delta E_S = g^* \mu_B B$ using $g^*$ from Eq.~\ref{gfactor} and for
$\Delta E_L = \hbar (e/m^*)B$.}}
\label{energy-split}
\end{figure}

Valuable insight into the relative importance of the various
effects can be obtained from a simple model, which takes the
Rashba effect and $sp-d$ exchange interaction into consideration,
but neglects bulk inversion asymmetry and nonparabolicity. With
this model the total level splitting energy for high Landau
numbers can be expressed to a first approximation as a function of
magnetic field according to;\cite{Das90,Pfeffer}
\begin{equation}
\label{delta}
\delta \approx \bigg [(\hbar \omega_c - g^* \mu_B B)^2 +
\Delta_R^2 \bigg]^{1/2} - \hbar \omega_c
\end{equation}
where $\Delta_R$ is the Rashba spin-orbit splitting energy.
In spite of their strong nonparabolic band structure, experimental and
theoretical values of $\hbar \omega_c$ for HgTe QWs are nearly
linear with magnetic field.\cite{Pfeuffer,Truchsess}
Using the theoretical value of  $\Delta_R$ and
the values of the effective $g$ factor according to Eq.~\ref{gfactor},
the total spin splitting energy, $\delta$, has been calculated 
for three gate voltages and is
plotted in Fig.~\ref{energy-split} together with $g^* \mu_B B$ and
$\hbar \omega_c$.
At high magnetic fields, the exchange
interaction tends to saturate and the Zeeman splitting,
i.e., $g_0\mu_B B$, corresponds to the value for a HgTe QW without Mn.
The experimental spin splitting energies from Eqs.~\ref{cos} and
\ref{nu} for three gate voltages are also plotted in 
Fig.~\ref{energy-split} versus $\hbar\omega_c$.
Obviously these values are in good agreement with the calculated
values of $\delta(B)$.


In conclusion, the spin transport properties of $n$ type \mmt{}/\cmt{}
QWs have been investigated.
The Rashba effect and giant Zeeman spin splitting have been separatly
investigated by varing the structure inversion asymmetry via a gate
voltage and by changing the temperature of the Mn ions, respectively.
We have experimentally demonstrated that the
Rashba effect in the magnetic 2DEG can be effectively controlled by means of
the structure inversion asymmetry via the gate voltage. These
results may be useful in realizing and
designing future spintronic devices.

The financial support of the Deutsche
Forschungsgemeinschaft (SFB 410), BMBF, and the DARPA SPINS program is
gratefully acknowledged.
One of us (Y.S.G.) would like to thank C. M. Hu for helpful
discussions and acknowledges the support of the Max Planck
Gesellschaft.

\footnotesize
\vspace{2mm}

$^*$ Permanent address: National Laboratory for Infrared
Physics, Shanghai Institute of Technical Physics, Chinese Academy of
Sciences, Shanghai 200083, China.


\begin{thebibliography}{99}
\bibitem{E2}For some gate voltages, additional population of the
second ($E2$) conduction band is observed. This does not, however,
influence the analysis of the $H1$ level splitting discussed here.
\end{thebibliography}
\end{document}